\documentclass[preprint,12pt]{elsarticle}

\usepackage{amssymb}
\usepackage{float}
\usepackage{xcolor}

\usepackage{amsmath}
\usepackage{physics}
\usepackage{subcaption}
\usepackage{graphicx}
\usepackage{amsthm}
\usepackage{dsfont}
\usepackage{hyperref} 

\journal{Chaos, Solitons \& Fractals}

\begin{document}

\begin{frontmatter}



\title{Charging free fermion quantum batteries}


\author[DIFI,CNR]{Riccardo Grazi} 
\ead{riccardo.grazi@edu.unige.it}
\affiliation[DIFI]{organization={Dipartimento di Fisica, Università di Genova},
            addressline={Via Dodecaneso 33}, 
            city={Genova},
            postcode={16146}, 
            country={Italy}}

\author[DIFI,CNR]{Fabio Cavaliere} 

\affiliation[CNR]{organization={CNR-SPIN},
            addressline={Via Dodecaneso 33}, 
            city={Genova},
            postcode={16146}, 
            country={Italy}
            }

\author[DIFI,CNR]{Maura Sassetti}
\author[DIFI,CNR]{Dario Ferraro}
\author[DIFI,CNR]{Niccolò Traverso Ziani}
\begin{abstract}
The performances of many-body quantum batteries strongly depend on the Hamiltonian of the battery, the initial state, and the charging protocol. In this article we derive an analytical expression for the energy stored via a double sudden quantum quench in \textcolor{black}{a large class of} quantum systems whose Hamiltonians can be reduced to 2x2 free fermion problems, whose initial state is thermal. \textcolor{black}{Our results apply to conventional two-band electronic systems across all dimensions} and quantum spin chains that can be solved \textcolor{black}{through} the Jordan-Wigner transformation. \textcolor{black}{In particular, we} apply our analytical relation to the quantum Ising chain, to the quantum XY chain, \textcolor{black}{to the cluster Ising and to the} long range SSH models. We obtain several results: (i) The strong dependence of the stored energy on the quantum phase diagram of the charging Hamiltonian persists even when the charging starts from a thermal state. Interestingly, in the thermodynamic limit, such a strong dependence manifests itself as non-analyticities of the stored energy corresponding to the quantum phase transition points of the charging Hamiltonian.
(ii) The dependence of the stored energy on the parameters of the Hamiltonian can, in the Ising chain case, be drastically reduced by increasing temperature; (iii) Charging the Ising or the XY chain prepared in the ground state of their classical points leads to an amount of stored energy that, within a large parameter range, does not depend on the charging parameters; (iv) The cluster Ising model and the long range SSH model, despite showing quantum phase transitions (QPTs) between states with orders dominated by different interaction ranges, do not \textcolor{black}{exhibit super-extensive, {\it i.e.} more than linear in the number of sites, scaling of the charging power}. 

\end{abstract}

\begin{graphicalabstract}
\end{graphicalabstract}

\begin{highlights}
\item Integrable systems are good candidates to be used as quantum batteries
\item Charging process dependence on quantum phase diagram from an initial thermal state
\item High temperatures can lead to energy plateaux using specific charging protocols
\item The enhancement of stored energy persists even in presence of long-range interactions
\end{highlights}


\end{frontmatter}

\section{Introduction}
The progressive miniaturization of solid-state devices opened the way to the multifaceted and fast developing field of quantum technologies, where the counterintuitive rules of quantum mechanics, instead of being an insurmountable limit for further developments, turned into a boost~\cite{Acin18, Zhang19, Raymer19, Sussman19}. In this context the realization of the first quantum computers represented a major step forward~\cite{Benenti_book}. This trend recently extended also towards energy storage, leading to the emergence of the idea of quantum batteries (QBs)~\cite{Alicki_2013}: systems devoted to store and transfer energy, working according to purely quantum mechanical effects instead of conventional electrochemistry~\cite{Bhattacharjee21, Quach23, Campaioli23}.

After a decade of theoretical proposals~\cite{Binder_2015, Campaioli_2017, Ferraro_2018, Andolina18, Barra_prl2019, Eckhardt2022, Rosa20, Seah21, Carrega16, Shaghaghi22, Cavaliere2023, Gyhm22, Morrone2023prapp, Mazzoncini23, Cavaliere24, Downing23, Gyhm24, Ahmadi24, Razzoli24, Pirmoradian19, Kamin20}, in the last few years the first experimental proofs of principle in this domain are appearing~\cite{Quach_2022, Hu_2022, Joshi22, Wenniger23, Gemme24}. The majority of them are based on collections of natural or artificial two-level systems (qubits), possibly interacting among them, promoted from the ground to an highly energetic many-body state by properly turning on and off the interaction with an external system playing the role of a charger~\cite{Crescente22}. 

Among the various systems considered as platforms for QBs, a great attention is currently devoted to the study of quantum spin chains~\cite{Le_2018, Dou22heis, Gemme23, Catalano23,Verma24, Yang24, Ali24}. The investigation of their charging dynamics can be increasingly demanding from the computational point of view by increasing the number of qubits composing them~\cite{Zhou20}. However, for peculiar choices of the coupling among the qubits, this problem can be overcome through a proper mapping of the spins into free fermions via the Jordan-Wigner transformation~\cite{Jordan28}. Therefore, limited to this class of integrable models~\cite{Franchini17}, the dynamics can be treated exactly allowing to explore the thermodynamical limit and leading to a complete control of the phase diagram of the system~\cite{Iyer13}. This is particularly remarkable for QBs and more in general quantum thermal machines due to the fact that, as recently observed, charging protocols crossing phase boundaries can play a major role in improving the efficiency and the stability of the energy storage~\cite{Campisi16, Barra_njp, Grazi24, Nello24, Grazi25}. 

The present work aims at providing a general framework to characterize charging protocols, seen for the sake of simplicity as a double sudden quench of one relevant parameter of the system, for this class of integrable QBs \cite{Banerjee24}. This allows us to determine the energy stored in the thermodynamic limit, as well as the averaged charging power, even for initial thermal states of the QBs which are typically quite difficult to approach using other techniques. Moreover, for charging protocols crossing peculiar phase boundaries, we have clear signatures of great robustness of the energy storage against changes in the Hamiltonian parameters, with important impact for technological applications. We have applied our general method to various relevant systems such as: the quantum Ising
chain~\cite{Suzuki12, Calabrese11}, the quantum XY chain~\cite{Lieb1961, Porta20, Sacco24}, the cluster Ising~\cite{Smacchia11, Ding19} and the long range SSH model~\cite{Gonzalez18}. The latter is not a spin chain model, but can be solved within the same proposed framework, further strengthening its generality. For each of the considered models we have highlighted the more interesting features in a QBs perspective.

\section{General Model}
\subsection{Non-superconducting systems}
The Hamiltonian of a generic time dependent two-band, non-interacting fermionic model defined on a periodic lattice, with periodic boundary conditions, can be written as
\textcolor{black}{
\begin{equation}
H(t)=\sum_{\mathbf{k}\in BZ} (c^\dagger_{a,\mathbf{k}},c^\dagger_{b,\mathbf{k}})(d_0(\mathbf{k},t) I_{2x2}+\mathbf{d}(\mathbf{k},t)\cdot \boldsymbol{\sigma})(c_{a,\mathbf{k}},c_{b,\mathbf{k}})^T. \label{H_Gen_Non_Superconduttiva}
\end{equation}}
Here, $\mathbf{k}$ is the quasimomentum and $t$ the time variable, $BZ$ is the Brillouin zone, $d_0(\mathbf{k},t)$ and $\mathbf{d}(\mathbf{k},t)=({d}_1(\mathbf{k},t),{d}_2(\mathbf{k},t),{d}_3(\mathbf{k},t))$ are free parameters, $I_{2x2}$ and $\boldsymbol{\sigma}$ are the identity and the Pauli matrix vector in the usual representation respectively. Finally $c_{\nu,\mathbf{k}}$, with \textcolor{black}{$\nu=a,b$}, is the fermionic annihilation operator for a fermion in the quantum state labelled by $\nu$ and by the quasimomentum $k$. As for the time dependence of the parameters, we restrict our attention to a double sudden quantum quench of duration $\tau$ with the initial and final parameters coinciding. Explicitly, we set ($j=0,..,3$)
\begin{equation}
d_j(\mathbf{k},t)=d^A_j(\mathbf{k})(\theta(-t)+\theta(t-\tau))+d^B_j(\mathbf{k})\theta(t)\theta(\tau-t).
\end{equation}
Here, $\theta(\cdot)$ is the Heaviside step function. We impose that the system at $t=0^-$ is described by a thermal density matrix with temperature $T$ parametrized by $\beta=1/(k_BT)$ ($k_B$ the Boltzmann constant), and that afterwards it evolves unitarily. The quantity we address is the energy $\Delta E(\tau)$ stored in the system immediately after the second quench, at $t= \tau + 0^+ = \tau^+$, namely
\begin{equation}
\Delta E(\tau)=\mathrm{Tr}\left\{\frac{e^{-\beta H(0^-)}}{\mathrm{Tr}\left\{{e^{-\beta H(0^-)}}\right\}}(H_H(\tau^+)-H_H(0^-))\right\},
\label{eq:res}
\end{equation}
where $\mathrm{Tr}\left\{\cdot\right\}$ represents the trace and $H_H(t)$ is the Hamiltonian in the Heisenberg representation. The explicit calculation is lengthy but straightforward. \textcolor{black}{Introducing} the two energy dispersions
\begin{eqnarray}
\epsilon(\mathbf{k})&=&d^A_0(\mathbf{k}) + \sqrt{\left(d^A_1(\mathbf{k})\right)^2+\left(d^A_2(\mathbf{k})\right)^2+\left(d^A_3(\mathbf{k})\right)^2} \label{epsilon_non_sc}\\
\omega(\mathbf{k})&=&d^B_0(\mathbf{k}) +\sqrt{\left(d^B_1(\mathbf{k})\right)^2+\left(d^B_2(\mathbf{k})\right)^2+\left(d^B_3(\mathbf{k})\right)^2},
\end{eqnarray}
we get
\begin{equation}
\Delta E(\tau)=\sum_{\mathbf{k}\in BZ}\frac{1-\cos(2\omega(\mathbf{k})\tau)}{\omega(\mathbf{k})^2\epsilon(\mathbf{k})}F_0(\mathbf{k})F_T(\mathbf{k},T,\mu). \label{FormulaGenerale_NON_SC}
\end{equation}
Here
\begin{eqnarray}
F_0(\mathbf{k})&=&\frac{\omega(\mathbf{k})^2}{\omega(\mathbf{k})^2-\left(d^B_3(\mathbf{k})\right)^2}\left(d^A_1(\mathbf{k})d^B_2(\mathbf{k})-d^A_2(\mathbf{k})d^B_1(\mathbf{k})\right)^2 \nonumber\\ &+& \bigg( d^A_3(\mathbf{k}) \sqrt{\omega(\mathbf{k})^2-\left( d^B_3(\mathbf{k})\right)^2}\nonumber\\ &-&\frac{d^B_3(\mathbf{k})}{\sqrt{\omega(\mathbf{k})^2-\left( d^B_3(\mathbf{k})\right)^2}}\left(d^A_1(\mathbf{k})d^B_1(\mathbf{k})+d^A_2(\mathbf{k})d^B_2(\mathbf{k})\right)\bigg)^2\nonumber\\
\, \label{F0}
\end{eqnarray}
and 
\begin{equation}
F_T(\mathbf{k},T,\mu)=n_F(d^A_0(\mathbf{k})-\epsilon(\mathbf{k}))-n_F(d^A_0(\mathbf{k})+\epsilon(\mathbf{k})), \label{FT}
\end{equation}
with $n_F(x)=1/(1+\exp(\beta(x-\mu)))$, the Fermi distribution with chemical potential $\mu$.\\
Several comments are in order. First, the term $(1-\cos(2\omega(\mathbf{k})\tau))$ encodes the time dependence of the stored energy on the charging time $\tau$. It clearly originates from having a collection of two-level systems and generically implies that, \textcolor{black}{for small $\tau$, one could expect} an oscillating behavior, followed by a plateau arising from summing over the different frequencies, and finally a recurrence of the oscillations if the quasimomentum is kept as a discrete variable. This kind of behavior was recently addressed in \cite{Grazi24}. The denominator $\omega^2(\mathbf{k})\epsilon(\mathbf{k})$ indicates a strong sensitivity of the stored energy on the level crossings of both the pre-quench Hamiltonian (via $\epsilon$), and even more drastically, of the Hamiltonian under which the system evolves (via $\omega^2$). Note that such strong dependence does not cause divergencies, since one finds that the numerator also goes to zero fast enough at the crossings. However, it can generate kinks when the charging is considered as a function of the parameters of the Hamiltonian. In particular, when $d^B_0(\mathbf{k})$=0, it allows for the detection of quantum phase transitions of the Hamiltonian describing the system for $0<t<\tau$. The function $F_0(\mathbf{k})$ ensures that, in the absence of time dependence of the Hamiltonian, no energy transfer happens. Finally, the thermal and chemical potential weight $F_T(\mathbf{k},T,\mu)$ accounts for two facts: the thermal smearing of the initial density matrix of the system at finite temperature, and the fact that no energy can be transferred to the couple of states with momentum $\mathbf{k}$ if they are both fully occupied or unoccupied. This happens because we perform a global quench on non interacting systems and the time evolution conserves the momentum of each quasiparticle. \textcolor{black}{It is also worth mentioning that all the stored energy can be extracted by unitary operations since our charging protocol is unitary.}

\subsection{Superconducting systems}
A simple although useful implication of the formula reported in Eq.(\ref{eq:res}) is that, through a canonical particle-hole transformation, it allows us to solve the same 'charging' problem in the case of a single species of fermions with superconducting correlations, and hence in the case of spin chains that can be exactly solved by the Jordan-Wigner transformation. Indeed, we can retrace what we have just discussed almost \textit{verbatim}. The time dependent Hamiltonian is given by
\begin{equation}
H_S(t)=\frac{1}{2}\sum_{\mathbf{k}\in BZ} (c^\dagger_{\mathbf{k}},c_{-\mathbf{k}})\left(Z(\mathbf{k},t)\sigma_z+X(\mathbf{k},t)\sigma_x\right)(c_{\mathbf{k}},c^\dag_{-\mathbf{k}})^T. \label{H_Gen_SC}
\end{equation}
Here, $c_{\mathbf{k}}$ is the spinless fermionic operator, and the terms proportional to the identity matrix have been set to zero since the system possesses a synthetic particle-hole symmetry \cite{de2018superconductivity}. Moreover, for the sake of simplicity, we have set to zero the coefficient of the $\sigma_y$ Pauli matrix. This can be safely done in all models of interest in the following by inserting the appropriate phase to the Fourier transformation~\cite{Franchini17}. The time dependence of the parameters is
\textcolor{black}{
\begin{eqnarray}
X(\mathbf{k},t)&=&X_A(\mathbf{k})(\theta(-t)+\theta(t-\tau))+X_B(\mathbf{k})\theta(t)\theta(\tau-t),\\
Z(\mathbf{k},t)&=&Z_A(\mathbf{k})(\theta(-t)+\theta(t-\tau))+Z_B(\mathbf{k})\theta(t)\theta(\tau-t).
\end{eqnarray}}
If we now set
\textcolor{black}{
\begin{eqnarray}
\epsilon_S(\mathbf{k})&=&\sqrt{X_A(\mathbf{k})^2+Z_A(\mathbf{k})^2},\label{ep}\\
\omega_S(\mathbf{k})&=&\sqrt{X_B(\mathbf{k})^2+Z_B(\mathbf{k})^2}\label{om},
\end{eqnarray}}
then we find for the energy stored
\textcolor{black}{
\begin{equation}
\Delta E_S(\tau)=\sum_{\mathbf{k}\in BZ}\frac{1-\cos(2\omega_S(\mathbf{k})\tau)}{2\epsilon_S(\mathbf{k})\omega_S^2(\mathbf{k})}\left(X_A(\mathbf{k})Z_B(\mathbf{k})-Z_A(\mathbf{k})X_B(\mathbf{k})\right)^2\tanh\left(\frac{\beta\epsilon_S(\mathbf{k})}{2}\right). \label{FormulaGenerale_SC}
\end{equation}}

\section{Applications}
In this section, we apply the previously derived relations to the quantum Ising chain, the XY chain, the cluster Ising chain and the extended SSH model to explore various applications. In all cases, we focus on the thermodynamic limit, $N \to \infty$. In this regime, the energy stored in the system as a function of the charging process duration $\tau$ initially exhibits pronounced oscillations. For larger values of $\tau$, these oscillations decrease in amplitude, leading the stored energy to approach a plateau. This plateau value, formally reached as $\tau \to \infty$, serves as a primary observable of interest throughout this article. For the cluster Ising model and the extended SSH model, we also analyze the maximum charging power, defined as~\cite{Andolina18}
\begin{equation} 
P_{max} = \max_{\tau}\left[\frac{\Delta E(\tau)}{\tau}\right]. \label{MaxPowerFormula}
\end{equation}
Furthermore, in the extended SSH model, we investigate the finite-size scenario. In this case, the energy stored as a function of $\tau$ reveals three distinct regimes instead of two, as previously reported in \cite{Grazi24, Rossini20}: these additional features arise due to recurrence effects linked to the finite size of the chain. Specifically, for long but finite chains, the energy stored resumes oscillating after the plateau. Our findings are as follows. 

For the Ising model, we analyze the plateau of the energy stored per site resulting from a quench of the external field. The plateau value exhibits a non-analytical dependence at the QPT of the model, even when the charging process begins from a thermal state. Additionally, we identify a parameter range where the plateau value is independent of the quench parameters.

For the XY model, by quenching the anisotropy parameter, we demonstrate that the plateau of the stored energy retains a non-analytical dependence, signaling the QPTs of the model. A region where the plateau value remains unaffected by the quench parameters is also identified.

For the cluster Ising model, we quench the two-spin interaction and observe QPT-related effects similar to those in the previous models. However, the maximum charging power, $P_{max}$, scales linearly with the number of sites, despite the presence of three-spin interactions.

Finally, for the extended SSH model, we quench the dimerization parameter, considering various types of nearest-neighbor interactions up to the third nearest neighbor. This approach validates the results even in a fermionic model with long-range interactions. For finite-sized chains, the stored energy oscillations as a function of $\tau$—arising from finite-size effects after the plateau—exhibit maxima, with respect to the model parameters, that occur both at the QPTs and at points of complete dimerization. Furthermore, even in this long-range model, the charging power scales linearly with $N$.

In all the models that will be discussed from now on, energies are given in units of a multiplicative parameter $\mathcal{J}$ of the Hamiltonian, which has the dimensions of energy and is consistently set to $\mathcal{J} = 1$ in every case.

\subsection{The Ising chain}
\begin{figure}[H]
    \centering
    \begin{subfigure}{0.45\textwidth}
        \centering
        \includegraphics[width=\linewidth]{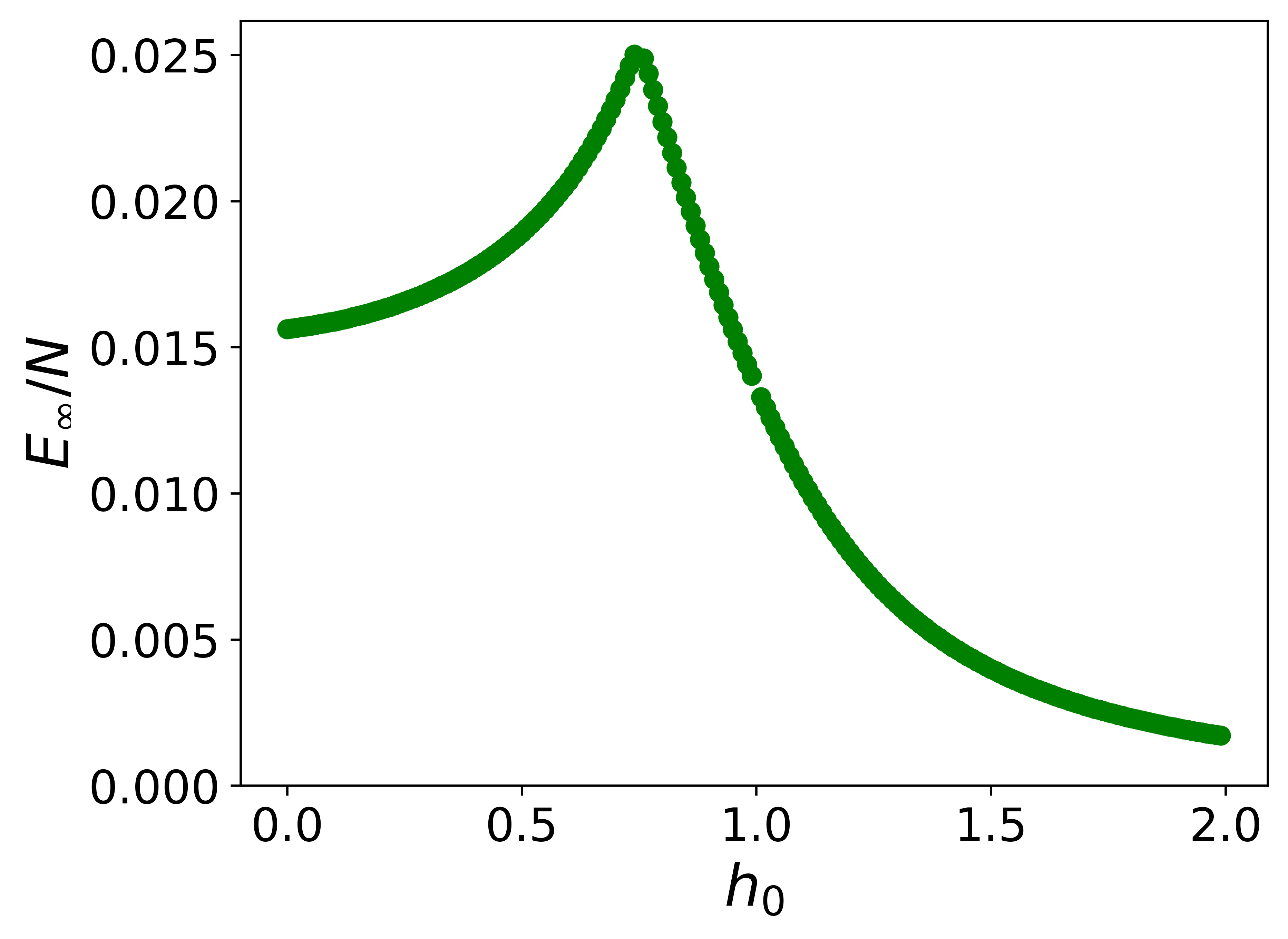}
        \caption{$h_1 = 0.25$ and $\beta = 10$}
        \label{Ising_Picco}
    \end{subfigure}
    \begin{subfigure}{0.45\textwidth}
        \centering
        \includegraphics[width=0.95\linewidth]{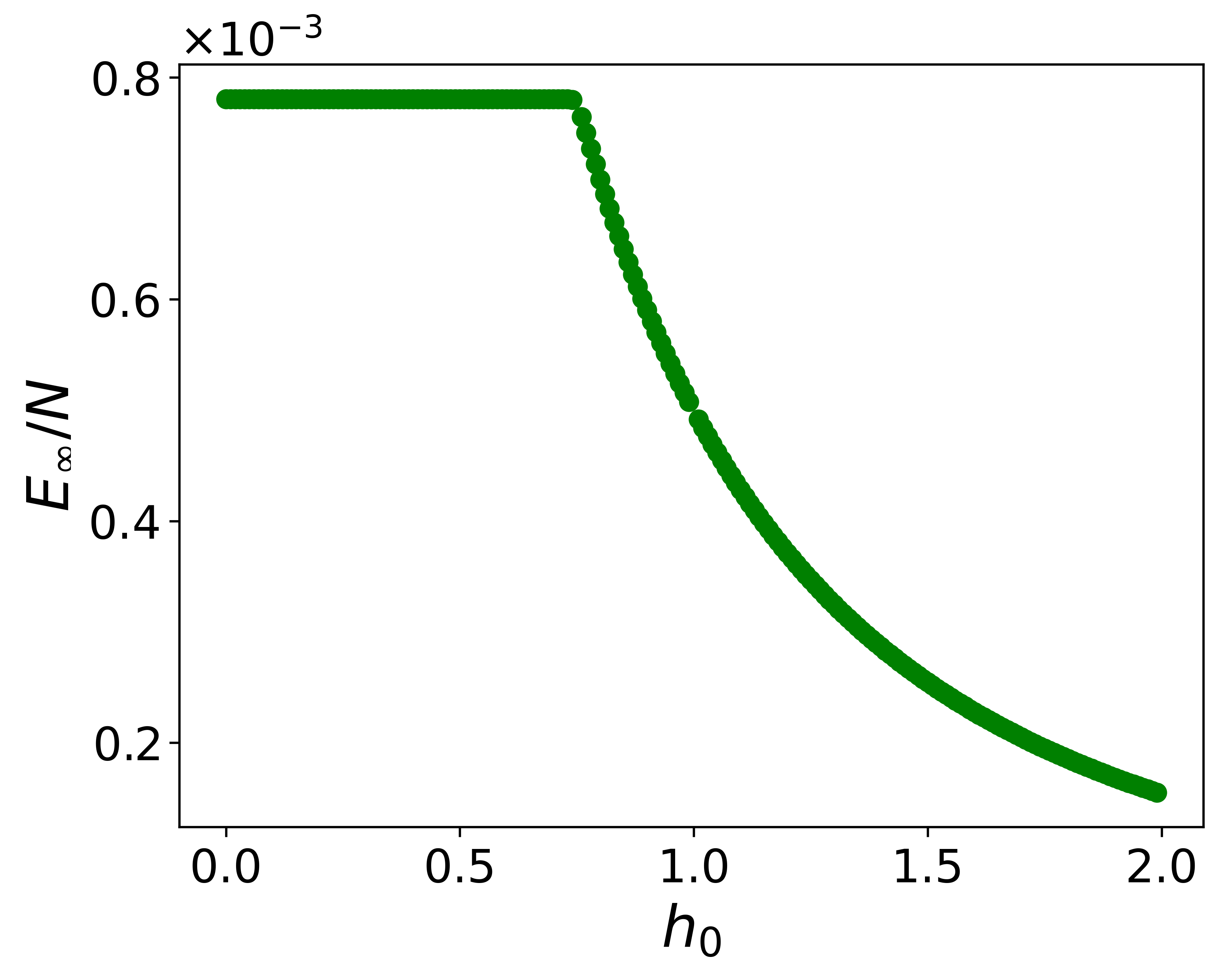}
        \caption{$h_1 = 0.25$ and $\beta = 0.1$}
        \label{Ising_Plateau}
    \end{subfigure}
    \caption{Energy stored per site in the $\tau \to \infty$ limit as a function of $h_0$ for an Ising-model-based QB in a \textcolor{black}{finite low-temperature} (a) and an high temperature (b) scenario.}
    \label{Ising}
\end{figure}

For this chain, we \textcolor{black}{initially operate a double quench by varying} the external field between the values $h_0$ and $h_0+h_1$. The dimensionless Hamiltonian of the quantum Ising chain in a transverse field reads~\cite{Suzuki12}
\begin{equation}
    H^{Ising} = -\frac{1}{2} \sum_{j=1}^N \left[\sigma_j^x \sigma_{j+1}^x+h \sigma_j^z\right].
\end{equation}
It is well known that this model presents a QPT at $h = 1$~\cite{Dziarmaga05}. Such a transition separates the ordered phase and the quantum paramagnetic one. A Jordan-Wigner mapping followed by a Fourier transformation can be used to write the Hamiltonian in terms of spinless fermionic operators $c_k$, leading to \cite{Franchini17}
\begin{equation}
H^{Ising} = \frac{1}{2} \sum_{{k}\in BZ}
\begin{pmatrix}
c_{k}^\dag & c_{-{k}}
\end{pmatrix}
\left[\left(h - \cos({k})\right)\sigma_z  -\sin({k}) \sigma_x\right]
\begin{pmatrix}
c_{k} & c_{- {k}}^\dag
\end{pmatrix}^T.\label{Ising_matrix}
\end{equation}
Here, the Brillouin zone is the interval $(-\pi,\pi)$ and the quasimomentum $k$ has only one component since we are in one dimension. The time-dependent problem we want to address is the one analyzed in the previous section. \textcolor{black}{Here the Hamiltonian reads}
\begin{equation}
H_S^{Ising}(t)=\frac{1}{2}\sum_{{k}\in BZ} (c^\dagger_{{k}},c_{-{k}})\left(Z^{Ising}({k},t)\sigma_z+X^{Ising}({k},t)\sigma_x\right)(c_{{k}},c^\dag_{-{k}})^T, \label{H_I_SC}
\end{equation}
with
\begin{equation}
    \begin{aligned}
        &X^{Ising}(k,t) = -\sin(k), \\
        &Z^{Ising}(k,t) = \left(h_0 - \cos(k)\right)(\theta(-t)+\theta(t-\tau))+\left(h_0 + h_1 - \cos(k)\right)\theta(t)\theta(\tau-t).
    \end{aligned}
\end{equation}
The dispersions $\epsilon_S^{Ising}(k)$ and $\omega_S^{Ising}(k)$ are readily obtained through Eqs.(\ref{ep})-(\ref{om}) respectively.\\
We aim at computing the energy  $\Delta E^{Ising}_S (\tau)$ stored in the system. Through Eq.(\ref{FormulaGenerale_SC}) we get
\begin{equation}
    \Delta E^{Ising}_S (\tau) = h_1^2 \sum_{k \in BZ}\frac{1-\cos(2\omega^{Ising}_S(k) \tau)}{2\epsilon^{Ising}_S(k)\left(\omega^{Ising}_S(k)\right)^2} \sin^2(k)\tanh\left(\frac{\beta\epsilon^{Ising}_S(k)}{2}\right). \label{EnergyStoredIsing}
\end{equation}
\textcolor{black}{It can be observed that the stored energy reported in Eq.\eqref{EnergyStoredIsing} is upper bounded by the maximum energy that the system can store, given by $\sum_k \epsilon^{Ising}_S(k)$ \cite{Hamma21}}. The result is represented in Fig.(\ref{Ising}). There, we have set $h_1=0.25$ and plotted the energy stored per site as a function of $h_0$ in the limit $\tau\rightarrow\infty$, indicated in the following as $E_{\infty}$, for $\beta = 10$ (panel (a)) and $\beta=0.1$ (panel (b)). Two physical effects are worth mentioning. First, see panel (a), the strong, non-analytical dependence of the stored energy for large $\tau$ found in correspondence of $h_0 + h_1 = 1$ in the thermodynamic limit $N\rightarrow\infty$, that is at the QPT of the Hamiltonian implementing the time evolution, persists even when the initial state is prepared at finite temperature. Then, see panel (b), we observe that as the temperature increases, the curve flattens more and more before the QPT, eventually forming a distinct plateau. This plateau could be significant from a technological perspective, as it creates a \textcolor{black}{very stable} region where one can freely choose from a \textcolor{black}{quite wide} range of parameters without affecting the stored energy. However, along with this technological advantage of temperature also comes a downside, as the maximum stored energy is reduced with increasing temperature.

\begin{figure}[H]
    \centering
    \begin{subfigure}{0.45\textwidth}
        \centering
        \includegraphics[width=\linewidth]{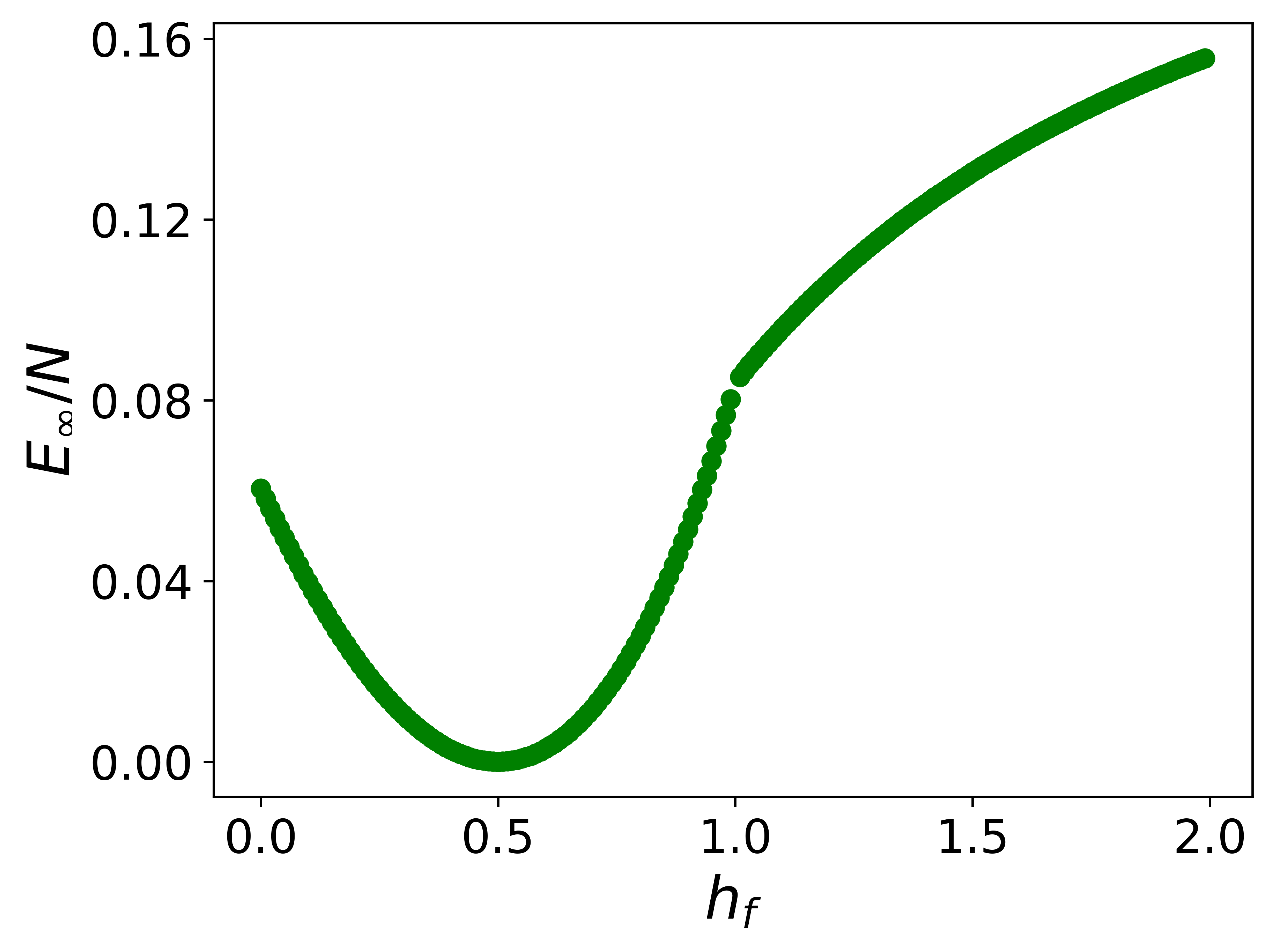}
        \caption{\textcolor{black}{$h_0 = 0.5$} and $\beta = 10$}
        \label{Ising_NoPlateau_Basso}
    \end{subfigure}
    \begin{subfigure}{0.45\textwidth}
        \centering
        \includegraphics[width=\linewidth]{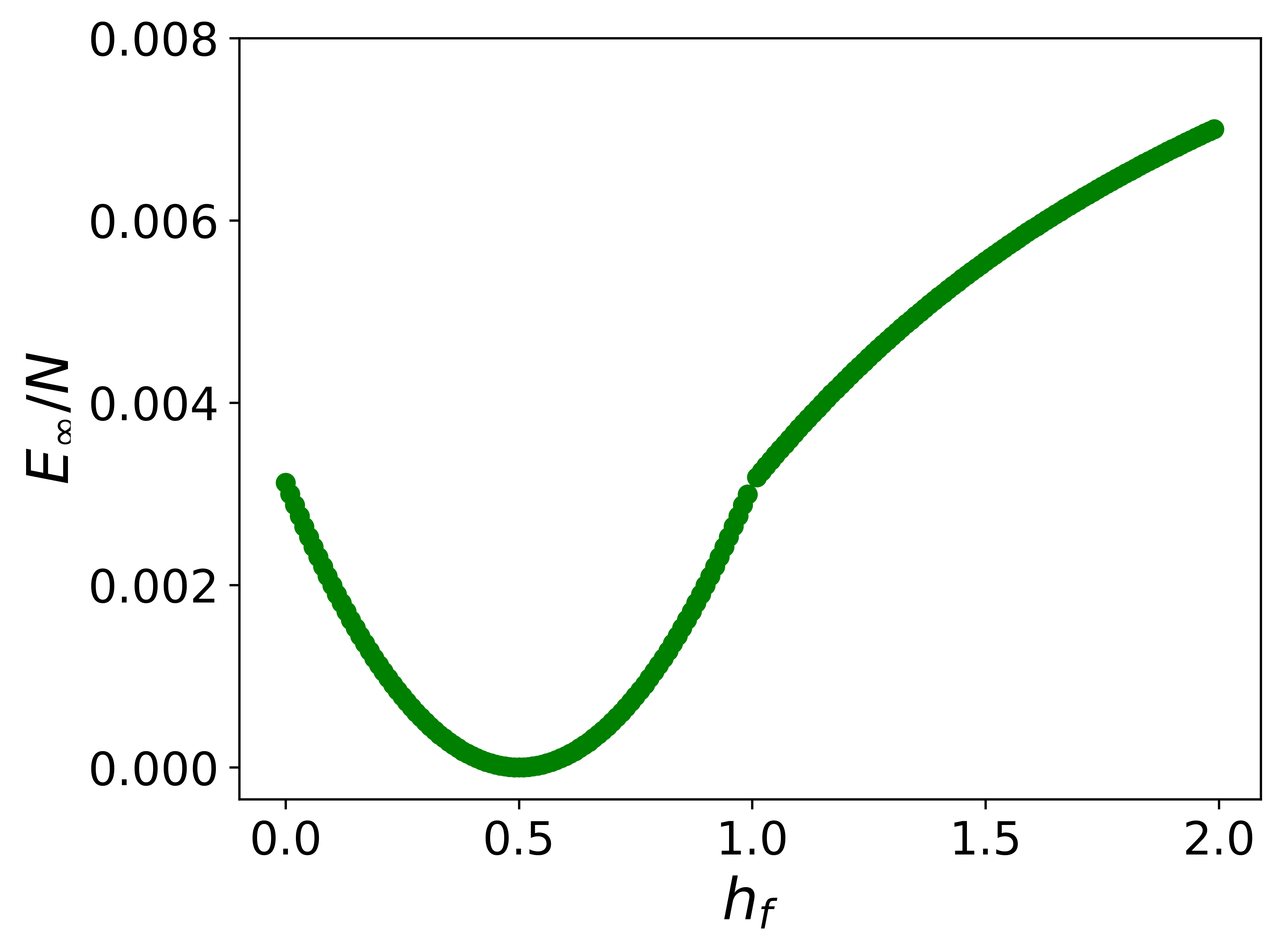}
        \caption{\textcolor{black}{$h_0 = 0.5$} and $\beta = 0.1$}
        \label{Ising_NoPlateau_Alto}
    \end{subfigure}
    \caption{Energy stored per site in the $\tau \to \infty$ limit as a function of $h_f$ for an Ising-model-based QB in a \textcolor{black}{finite low-temperature} (a) and an high temperature (b) scenario.}
    \label{Ising_Muovendo_H1}
\end{figure}

\begin{figure}[H]
    \centering
    \includegraphics[width=0.8\linewidth]{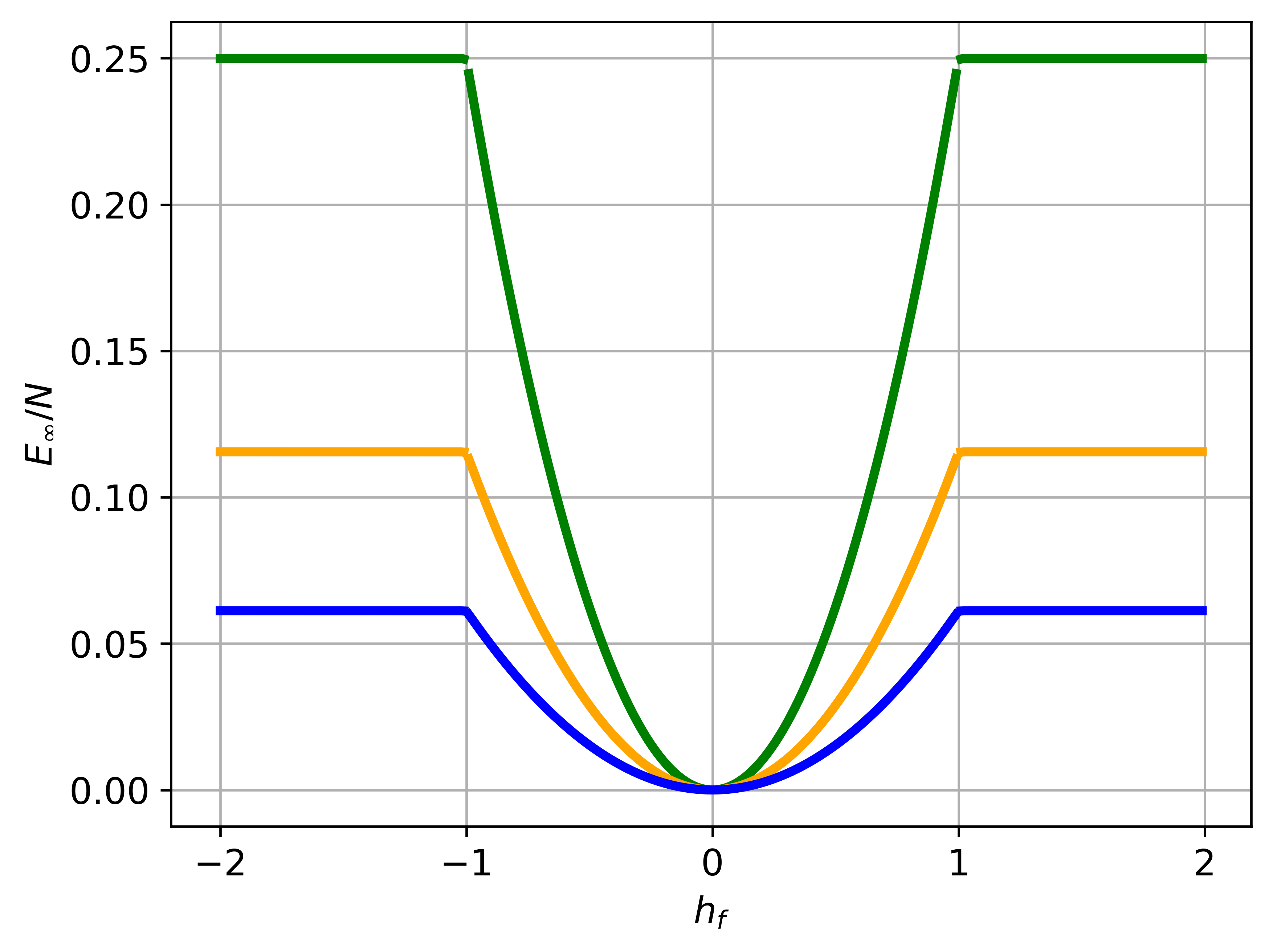}
    \caption{Energy stored per site in the $\tau \to \infty$ limit as a function of $h_f$ for an Ising-model-based QB with $N = 300$ sites starting from \textcolor{black}{$h_0 = 0$} at $\beta=0.5$ (blue curve), $\beta=1$ (orange curve) and $\beta \to \infty$ (green curve).}
    \label{Ising_Muovendo_H1_Plateau}
\end{figure}

\textcolor{black}{Let us consider a double quench, starting again from the parameter $h_0$, but now examining its behavior as a function of $h_f=h_{0}+h_{1}$. The goal of this change is to determine whether varying the field enables the observation of a plateau even at low temperatures.} In Fig. \ref{Ising_Muovendo_H1}, we see that for a generic value of \textcolor{black}{$h_0$}, even though the function remains non-analytical in the thermodynamic limit in correspondence of the QPT of the charging Hamiltonian, no plateau appears as we vary $h_f$, neither at low (Fig. \ref{Ising_NoPlateau_Basso}) nor at high (Fig. \ref{Ising_NoPlateau_Alto}) temperatures. The minimum that appears at $h_f = 0.5$ is justified by the fact that if \textcolor{black}{$h_0 = h_f$ ($h_{1}=0$)}, no quench takes place.
However, if we specifically start our evolution from the classical point \textcolor{black}{$h_0 = 0$} \cite{Maric20} and then we turn on an external field up until a final value $h_f$ we can see in Fig.~\ref{Ising_Muovendo_H1_Plateau} the appearance of a plateau after the critical value $h_f = 1$ also at zero temperature. The situation \textcolor{black}{occurs symmetrically also} for negative values of $h_f$.
The presence of the plateau can be demonstrated analytically. By setting \textcolor{black}{$h_0 = 0$} and $\beta \to \infty$, in the thermodynamic limit $N\rightarrow\infty$ we get
\begin{equation}
\begin{aligned}
    \Delta E^{Ising}_S (\tau \to \infty) = &\frac{N}{4 \pi} h_f^2 \int_{-\pi}^{+\pi}\frac{\sin^2(k)}{1 + h_f^2 - 2h_f \cos(k)} dk. \label{Integral_Ising}
\end{aligned}
\end{equation}
The integral can be analytically computed, and we obtain
\begin{equation}
    \Delta E^{Ising}_S (\tau \to \infty) = \begin{cases}
        \frac{N}{4} \quad &\text{if} ~|h_f| > 1 \\
        \frac{N}{4} h_f^2 \quad &\text{if} ~|h_f| < 1
    \end{cases}
\end{equation}
leading to the observed plateau even at zero temperature. We can also observe, from Fig.~\ref{Ising_Muovendo_H1_Plateau}, that the effect of temperature is just to decrease the value of those plateau.


\subsection{The XY chain}
\begin{figure}[H]
    \centering
    \includegraphics[width=\linewidth]{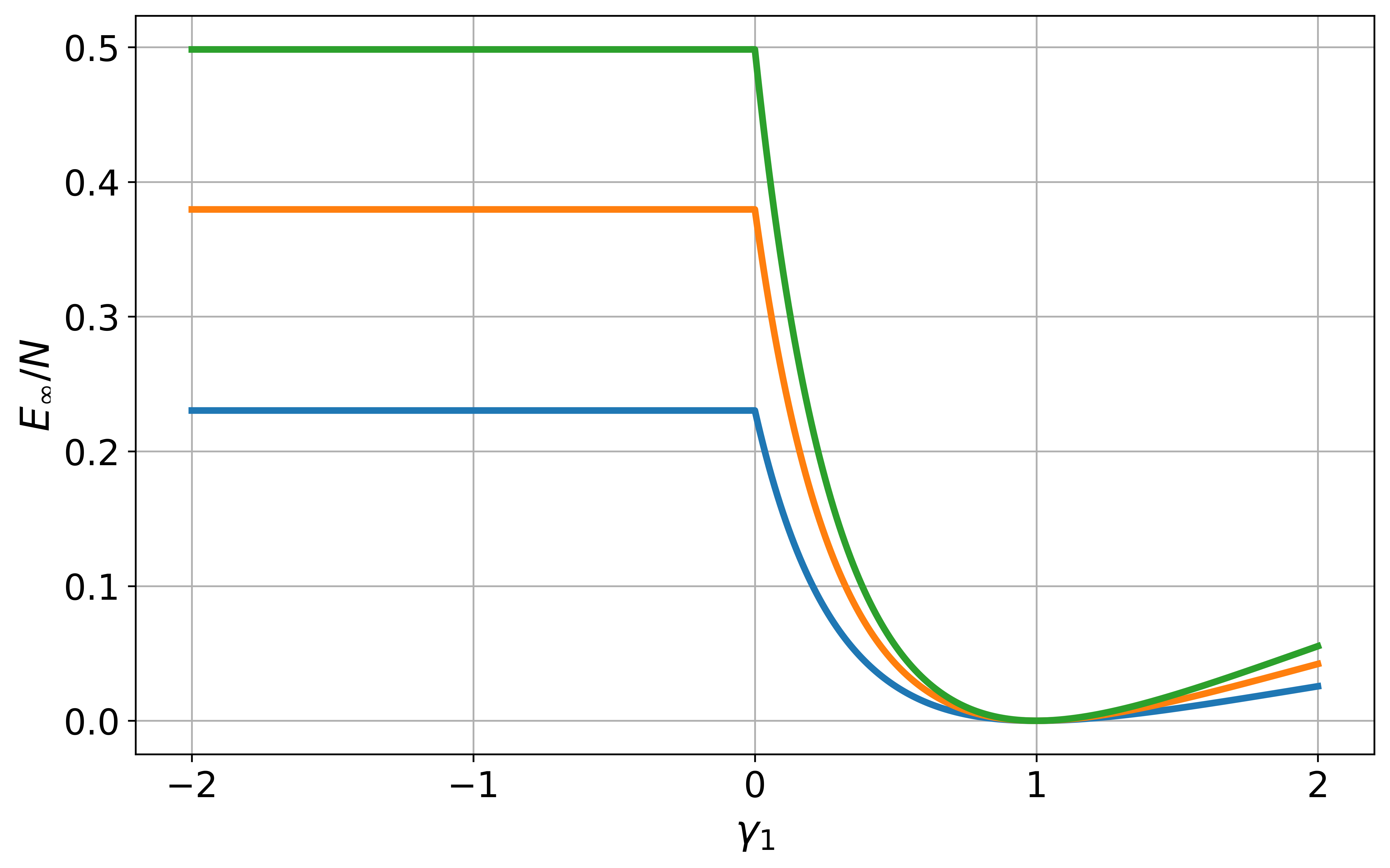}
    \caption{Energy stored per site in the $\tau \to \infty$ limit for an XY chain with $N = 300$ sites as a function of $\gamma_1$ for $\beta = 1$ (blue curve), $\beta = 2$ (orange curve) and $\beta \to \infty$ (green curve). All external fields are set to zero.}
    \label{XYPlot}
\end{figure}

We now address the XY chain, a system with a slightly larger parameter space with respect to the Ising chain~\cite{Lieb1961}, in order to find a \textcolor{black}{different} scenario \textcolor{black}{where} the zero-temperature plateau just described \textcolor{black}{also appears}. The dimensionless Hamiltonian of the quantum XY chain reads
\begin{equation}
    H^{XY} = -\frac{1}{2} \sum_{j=1}^N\left[\left(\frac{1+\gamma}{2}\right) \sigma_j^x \sigma_{j+1}^x+\left(\frac{1-\gamma}{2}\right) \sigma_j^y \sigma_{j+1}^y + h\sigma_j^z\right], \label{XY_Ham}
\end{equation}
where $\gamma$ is the anisotropy parameter. Just like the quantum Ising model, that can be recovered by setting $\gamma = 1$, this Hamiltonian can be re-written in the form of Eq.\eqref{H_Gen_SC} as follows
\begin{equation}
H^{XY} = \frac{1}{2} \sum_{{k}\in BZ}
\begin{pmatrix}
c_{k}^\dag & c_{-{k}}
\end{pmatrix}
\left[ \left(h - \cos({k})\right) \sigma_z  -\gamma \sin({k}) \sigma_x\right]
\begin{pmatrix}
c_{k} & c_{- {k}}^\dag
\end{pmatrix}^T.
\end{equation}
Here, two parameters can be quenched: the anisotropy parameter $\gamma$ and the external field $h$, such that
\begin{equation}
    \begin{aligned}
        &X^{XY}(k,t) = \left(-\gamma_0 \sin(k)\right)(\theta(-t)+\theta(t-\tau))+\left(-\gamma_1 \sin(k)\right)\theta(t)\theta(\tau-t) \\
        &Z^{XY}(k,t) = \left(h_0 - \cos(k)\right)(\theta(-t)+\theta(t-\tau))+\left(h_1 -\cos(k)\right)\theta(t)\theta(\tau-t).
    \end{aligned}
\end{equation}
Therefore, the time-dependent energy stored at zero temperature is:
\begin{equation}
    \Delta E^{XY}_S (\tau) = \sum_{k \in BZ}\frac{1-\cos(2\omega^{XY}_S(k) \tau)}{2\epsilon^{XY}_S(k)(\omega^{XY}_S(k))^2}  \sin^2(k) \left[\gamma_0(h_1 - \cos(k)) - \gamma_1(h_0 - \cos(k))\right]^2. \label{Discrete_XY}
\end{equation}
We now make two assumptions: the first, which is crucial for the emergence of the desired effect, is to start the evolution from the classical point of the model, $(\gamma_0, h_0) = (1,0)$, just like \textcolor{black}{in the case of} the Ising model of the previous section. The second assumption, made for the \textcolor{black}{sake of} simplicity but not essential, is to quench only the anisotropy parameter, keeping $h_0 = h_1 = 0$. Since we are interested in the thermodynamic limit, we can write Eq.\eqref{Discrete_XY} as an integral and, with the aforementioned conditions, obtain
\begin{equation}
    \Delta E^{XY}_S (\tau \to \infty) = - \frac{N}{2 \pi} (1 - \gamma_1)^2 \int_{0}^{\pi}\frac{\sin^4(k) - \sin^2(k)}{1 + (\gamma_1^2 - 1)\sin^2(k)} dk \label{integrale_xy}
\end{equation}
The integral is analytically solvable just like the one of Eq.\eqref{Integral_Ising}. The result is (see \ref{Appendix} for more details)
\begin{equation}
    \begin{cases}
         \frac{N}{4} \quad &\text{if} ~ \gamma_1 < 0\\
         \frac{N}{4} \frac{(\gamma_1 - 1)^2}{(\gamma_1 + 1)^2} \quad &\text{if} ~ \gamma_1 > 0. \label{Result_Integral_XY}
    \end{cases}
\end{equation}
So, as long as the evolution starts from $\gamma_0 = 1$, there is a region where the energy stored in the XY model in the thermodynamic limit remains independent of $\gamma_1$, i.e. the value of the anisotropy parameter after the quench, and this results in the formation of a plateau even at zero temperature. This plateau also appears for systems with a finite number of particles, as can be seen in Fig.~\ref{XYPlot}, where the energy stored per site has been plotted as a function of $\gamma_1$ for various values of $\beta$. Additionally, the effect of temperature is to reduce the value of the plateau by a factor $\tanh(\beta/2)$, just like observed in the previous section.

\subsection{The cluster Ising model}
\begin{figure}[H]
    \centering
    \begin{subfigure}{0.45\textwidth}
        \centering
        \includegraphics[width=0.95\linewidth]{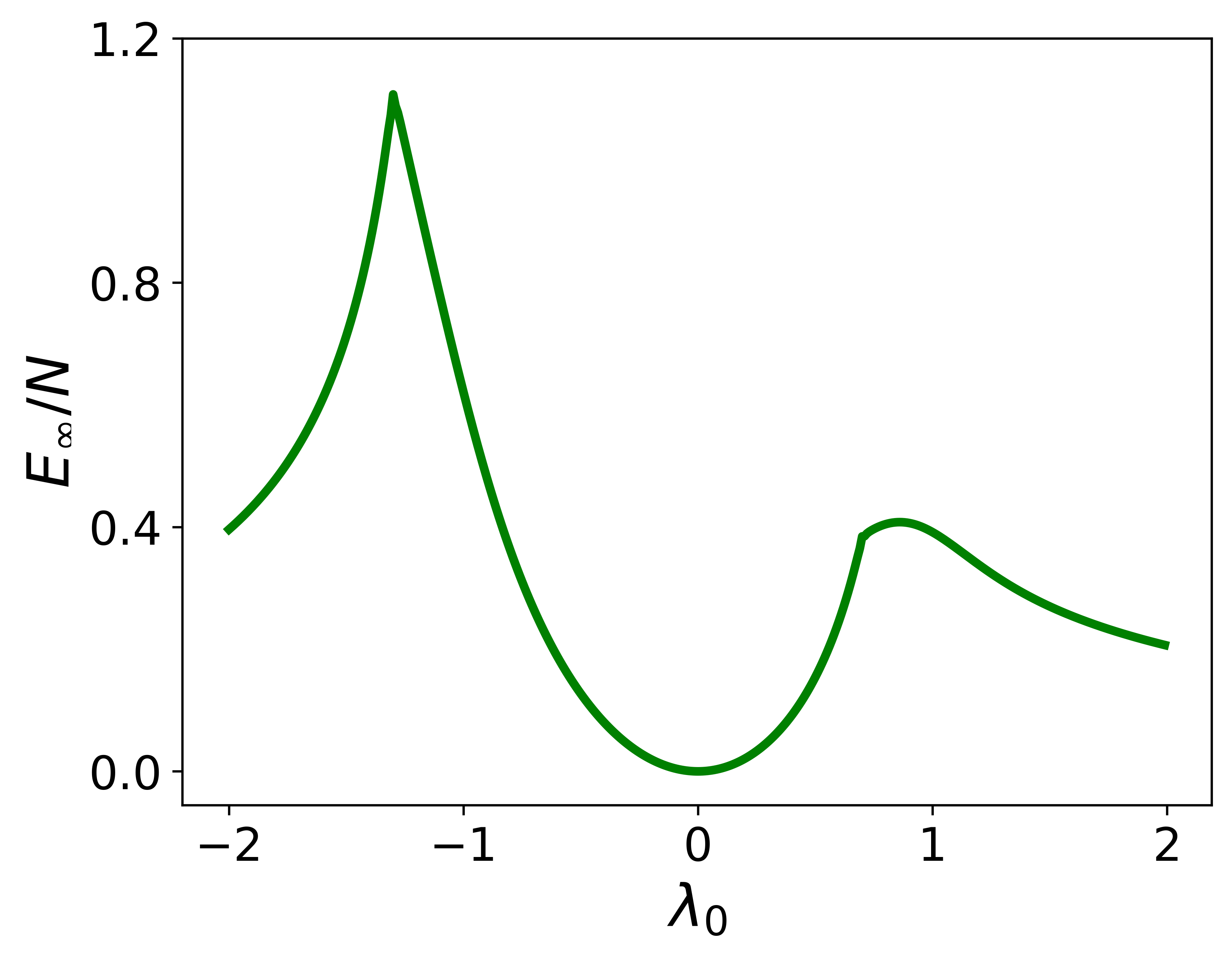}
        \caption{$\lambda_1 = 0.3$ and $\beta = 10$}
        \label{C-I_BasseT}
    \end{subfigure}
    \begin{subfigure}{0.45\textwidth}
        \centering
        \includegraphics[width=\linewidth]{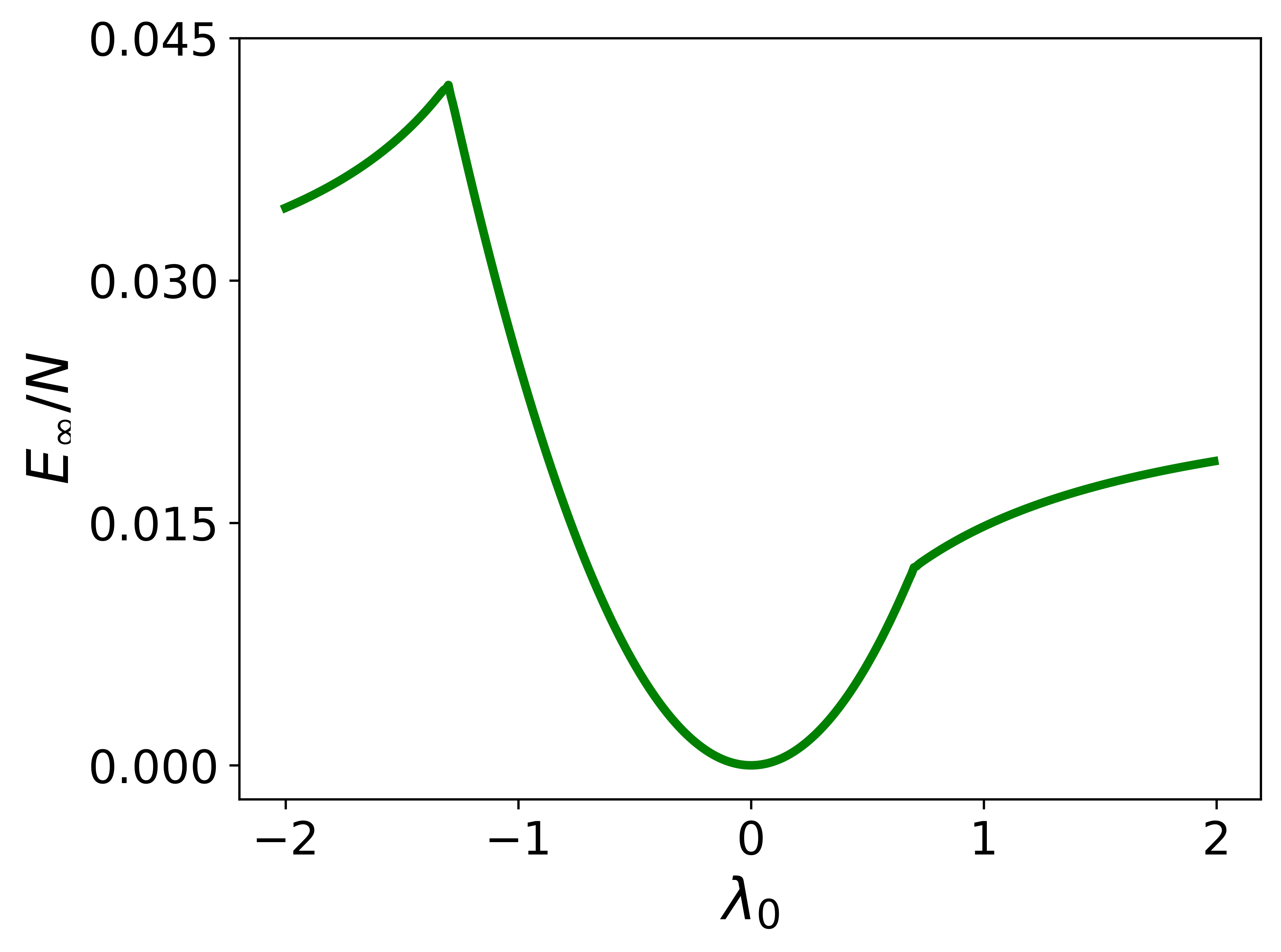}
        \caption{$\lambda_1 = 0.3$ and $\beta = 0.1$}
        \label{C-I_AlteT}
    \end{subfigure}
    \caption{Energy stored per site in the $\tau \to \infty$ limit as a function of $\lambda_0$ for $\lambda_1 = 0.3$ in a \textcolor{black}{finite low-temperature} (a) and an high temperature (b) scenario.}
    \label{Cluster_Ising_Plot}
\end{figure}

To clarify the role of long-range interactions, the last spin model we study is the cluster Ising model \cite{Smacchia11}, which exhibits \textcolor{black}{next-to-nearest-neighbour interactions due to the presence of a three-spin term that maps into a combination of hopping and superconducting terms at the fermionic level.} Its dimensionless Hamiltonian reads
\begin{equation}
    H_{C-I} = - \sum_{j = 1}^N \sigma_{j-1}^x \sigma_j^z \sigma_{j+1}^x + \lambda \sum_{j=1}^N \sigma_j^y \sigma_{j+1}^y.
\end{equation}
The fermionic Hamiltonian of the model in the momentum space is
\begin{equation}
H_S^{C-I}(t)=2\sum_{{k}\in BZ} (c^\dagger_{{k}},c_{-{k}})\left(Z^{C-I}({k},t)\sigma_z+X^{C-I}({k},t)\sigma_x\right)(c_{{k}},c^\dag_{-{k}})^T,
\end{equation}
with
\begin{equation}
    \begin{aligned}
        &X^{C-I}({k},t) = \sin(2k) + \lambda(t) \sin(k)\\
        &Z^{C-I}({k},t) = -\cos(2k) + \lambda(t) \cos(k)
    \end{aligned}
\end{equation}
where $\lambda(t)$ is the parameter that's going to be quenched as follows
\begin{equation}
    \lambda(t) = \lambda_0 (\theta(-t)+\theta(t-\tau)) + (\lambda_0 + \lambda_1) \theta(t)\theta(\tau-t).
\end{equation}
From Eq.\eqref{ep} we can observe that
\[
\epsilon(k) = 0 \implies \cos(3k) = \frac{1 + \lambda_0^2}{2\lambda_0}
\]
and since $|(1 + \lambda_0^2)/(2\lambda_0)| \geq 1$ for every value of $\lambda_0$, we have QPTs only when $\lambda_0 = 1$ (in $k = \frac{2}{3}n\pi$, $n \in \mathds{Z}$) and when $\lambda_0 = -1$ (in $k = \frac{2n+1}{3}\pi$, $n \in \mathds{Z}$), while setting $\lambda_0 = 0$ results in the flat bands configuration. The quantum phase transitions separate regimes where the two-spin physics dominates, and a clustered regime. The stored energy of the system reads
\begin{equation}
\Delta E_S^{C-I}(\tau)=\lambda_1^2\sum_{k\in BZ}\frac{1-\cos(2\omega^{C-I}_S(k)\tau)}{2\epsilon^{C-I}_S(k)(\omega^{C-I}_S(k))^2} \sin^2(3k)\tanh\left(\frac{\beta\epsilon^{C-I}_S(k)}{2}\right).
\end{equation}
Fig. \ref{Cluster_Ising_Plot} represents the energy stored per site for fixed $\lambda_1 = 0.3$ and also here we have the emergence of non-analyticities related to QPTs for $\lambda_0 = -1.3$ (such that $\lambda_0 + \lambda_1 = -1$) and $\lambda_0 = 0.7$ (such that $\lambda_0 + \lambda_1 = 1$) both in the low-temperature (panel (a)) and high-temperature (panel (b)) scenario.
However, it is possible to determine that the maximum charging power, defined in Eq.\eqref{MaxPowerFormula} \textcolor{black}{grows linearly with} the number of sites.


\subsection{SSH model}
An interesting, natively fermionic model whose Hamiltonian can be written in the form reported in Eq.\eqref{H_Gen_Non_Superconduttiva} is the SSH model. Considering \textcolor{black}{the general case of} $N^{th}$-neighbor interactions, this Hamiltonian takes the form \cite{Gonzalez18}
\begin{equation}
H_{SSH} = \sum_{|i-j| \leq N} J_{ij} (c_i^\dagger c_j + h.c.) \label{SSH_Ham_Iniziale}
\end{equation}
where dimensionless $J_{ij} = J_{ij}^*$ is the hopping amplitude between site \textit{i} and site \textit{j}. Since it is more convenient to think \textcolor{black}{about this problem} in terms of unit cells instead of single sites, we will identify from now on all the sites with odd indices as belonging to sublattice A and all the sites with even indices belonging to sublattice B, as follows 
\begin{equation}
    A_j \equiv c_{2j-1}, \quad B_j \equiv c_{2j}.
\end{equation}
\begin{figure}[H]
    \centering
    \includegraphics[width=0.8\linewidth]{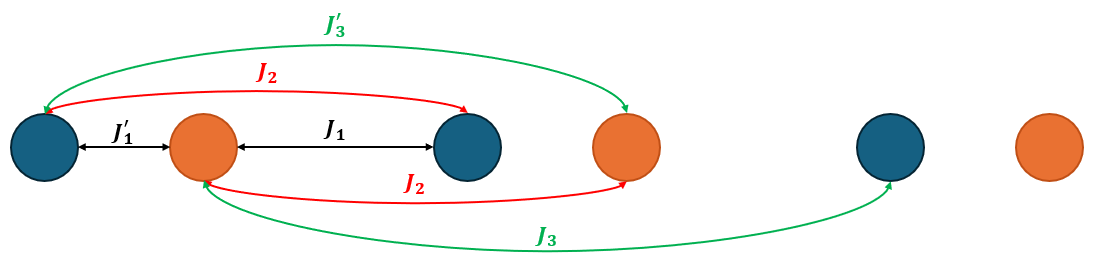}
    \caption{Representation of the SSH model with sketch of interactions up to third nearest-neighbor interactions. Blue circles represent sites belonging to sublattice A while orange circles represent sites belonging to sublattice B.}
    \label{SSH_Interactions}
\end{figure}
Another key feature to highlight is the fact that all even hoppings have the same length, regardless of the site from which they originate, while odd hoppings have a length that depends on the starting site. For example, if the distance between sites within the same unit cell is $\mathcal{D}$ and the distance between sites in adjacent cells is $\mathcal{L}$, then every second-neighbor hopping will cover a distance of $\mathcal{D} + \mathcal{L}$, no matter whether one starts from site A or B, but for third-neighbor hoppings, starting from site A in one unit cell, the hopping distance is $2\mathcal{D} + \mathcal{L}$, while starting from site B the distance becomes $\mathcal{D} + 2\mathcal{L}$. To distinguish those situations, the following notation can be used
\begin{equation*}
    \begin{cases}
        &J_{2j - n, 2j} \equiv J'_n \quad \text{and} \quad J_{2j, 2j + n} \equiv J_n \quad \text{for}~ n~ \text{odd} \\
        &J_{j, j \pm n} \equiv J_n \quad \text{for}~ n~ \text{even}
    \end{cases}
\end{equation*}
\noindent as can be seen in Fig. \ref{SSH_Interactions}. \textcolor{black}{In} momentum space we pass from $A_j$ and $B_j$ operators to \textcolor{black}{$c_{a,k}$ and $c_{b,k}$} ones, so that the Hamiltonian \eqref{SSH_Ham_Iniziale} can be rewritten in the form of Eq.\eqref{H_Gen_Non_Superconduttiva} with
\begin{equation}
\begin{aligned}
&d_0(k,t) = \sum_{p} 2 J_{2p}(t) \cos(pk), \\
&d_1(k,t) = \sum_p \left[J'_{2p-1}(t) \cos((p-1)k) + J_{2p-1}(t) \cos(pk)\right], \\
&d_2(k,t) = \sum_p \left[J_{2p-1}(t) \sin(pk) - J'_{2p-1}(t) \sin((p-1)k)\right], \\
&d_3(k,t) = 0,
\end{aligned} \label{d_vector_SSH}
\end{equation}
with $p$ ranging from 1 to $N/2$ if $N$ is even or $(N + 1)/2$ if $N$ is odd. Let us investigate the model starting from the nearest-neighbor scenario. 
\subsubsection{Nearest-neighbor interactions}
\begin{figure}[H]
    \centering
    \begin{subfigure}{0.45\textwidth}
        \centering
        \includegraphics[width=\linewidth]{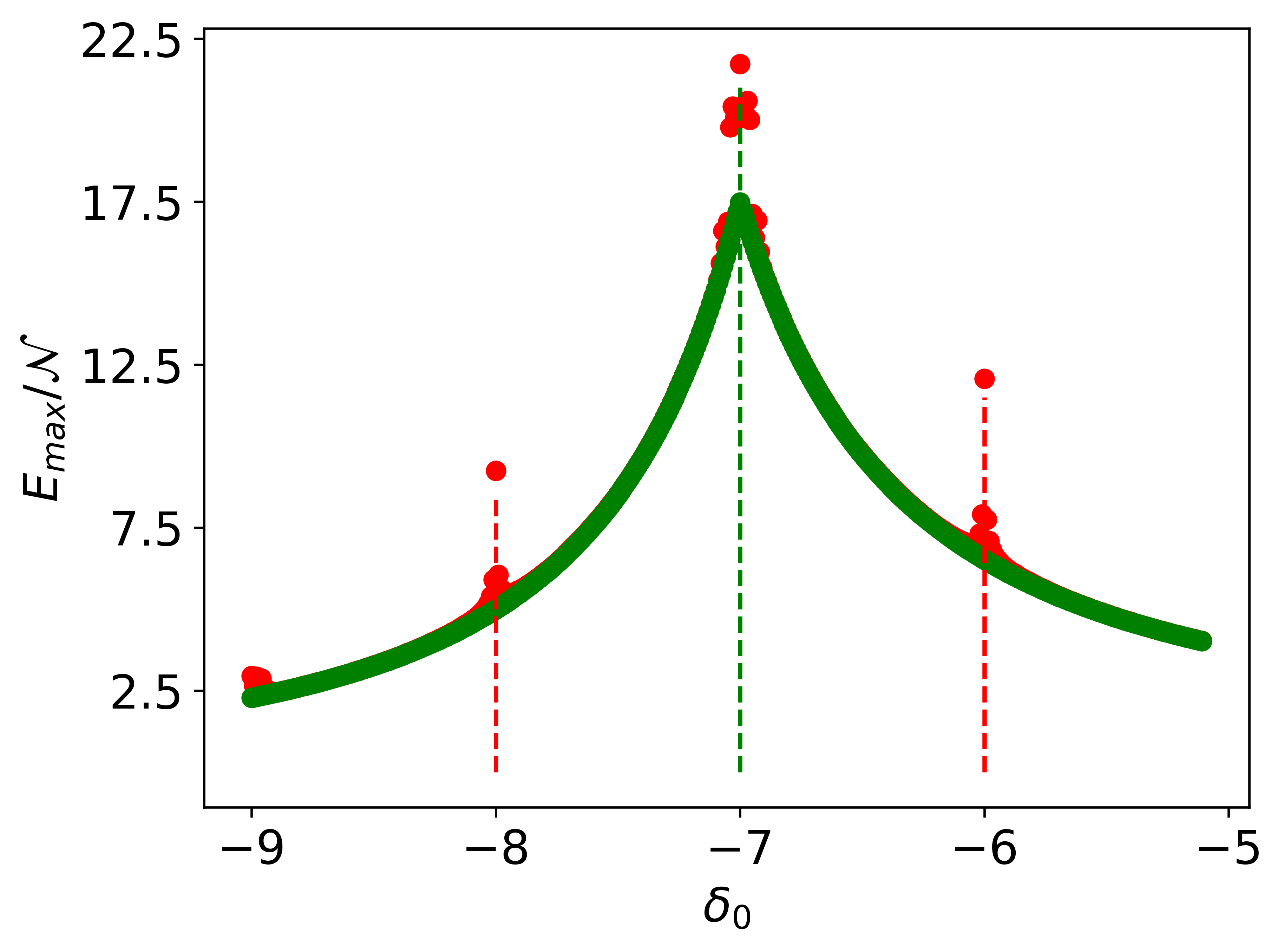}
    \caption{$\delta_1 = 7$ and $\beta = 10$}
    \label{SSH_NO_TEMP}
    \end{subfigure}
    \begin{subfigure}{0.45\textwidth}
        \centering
        \includegraphics[width=0.95\linewidth]{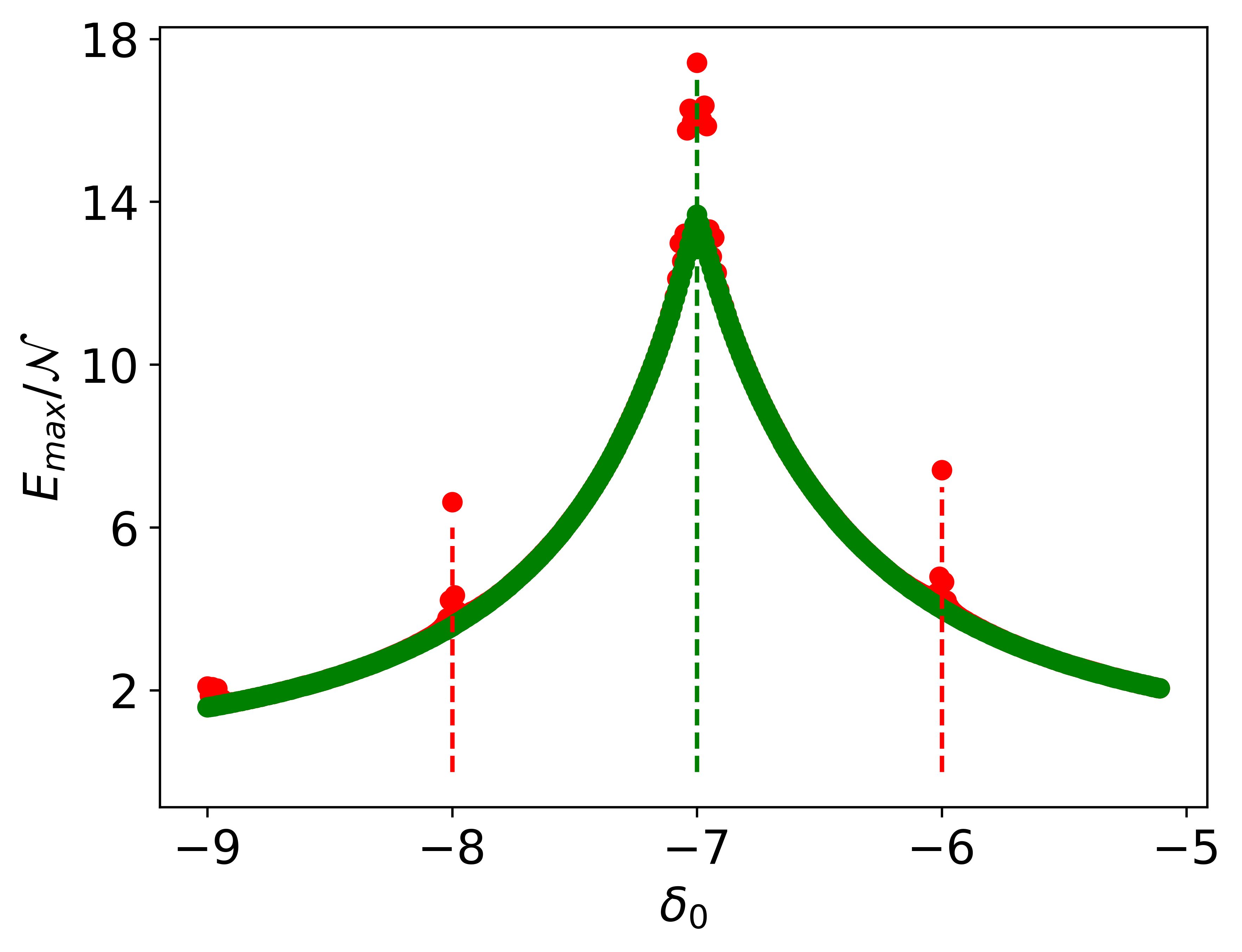}
        \caption{$\delta_1 = 7$ and $\beta = 0.1$}
        \label{SSH_TEMP}
    \end{subfigure}
    \caption{Maximum energy stored per dimer as a function of $\delta_0$ in correspondence of the recurrence time (red) and in the $\tau \to \infty$ limit (green) in a low-temperature (a) and high-temperature (b) scenario.}
    \label{SSH_Nearest}
\end{figure}

In this configuration, the fermionic Hamiltonian becomes
\textcolor{black}{
\begin{equation}
    H_{SSH}(t)=\sum_{k\in BZ} (c^\dagger_{a,k},c^\dagger_{b,k})\left[(1 + \delta + (1 - \delta) \cos(k)) \sigma_x + (1 - \delta) \sin(k) \sigma_y \right](c_{a,k},c_{b,k})^T
\end{equation}
}
where we set $J'_1 \equiv 1 + \delta$, $J_1 \equiv 1 - \delta$ and $J_{n \neq 1} = 0$, with $\delta \in \mathds{R}$ the dimerization parameter, which is the one that will be quenched. In this model, the QPT is topological, separating the trivial phase, characterized by having winding number \textcolor{black}{$w = 0$}, from the topological phase, with \textcolor{black}{$w = 1$}~\cite{Palyi2016-md}. Consistently with the results given in Eq.\eqref{d_vector_SSH} setting $p = 1$, the non-zero components of the $d-$vector are
\begin{equation}
\begin{aligned}
&d_1(k,t) = 1 + \delta(t) + (1 - \delta(t)) \cos(k), \\
&d_2(k,t) = (1 - \delta(t)) \sin(k) \\
\end{aligned}
\end{equation}
with
\begin{equation}
    \delta(t) = \delta_0 (\theta(-t)+\theta(t-\tau)) + (\delta_0 + \delta_1) \theta(t)\theta(\tau-t).
\end{equation}
From the form of $\epsilon(k)$, obtainable from Eq.\eqref{epsilon_non_sc}
\begin{equation}
    \epsilon(k) = \sqrt{2}\sqrt{1+\delta^2 + (1-\delta^2) \cos(k)}.
\end{equation}
We can observe that the system crosses a QPT line when
\begin{equation}
    \epsilon(k) = 0 \implies \cos(k) = \frac{1 + \delta^2}{1 - \delta^2} \geq 1.
\end{equation}
Therefore, the only \textcolor{black}{solution is} $\cos(k) = 1$, so $\delta = 0$. Another important observation is that, for $\delta = \pm 1$, the bands of the system become flat, i.e. independent of $k$, and the charging Hamiltonian \textcolor{black}{can be seen as a collection of} disconnected dimers. We can now compute Eq.\eqref{FormulaGenerale_NON_SC} to yield
\begin{equation}
\Delta E^{SSH}(\tau)=4\delta_1^2 \sum_{k\in BZ}\frac{1-\cos(2\omega^{SSH}(k)\tau)}{\epsilon^{SSH}(k)(\omega^{SSH}(k))^2} \sin^2(k) \tanh\left(\frac{\beta\epsilon^{SSH}(k)}{2}\right).
\end{equation}
As already seen in~\cite{Grazi24}, we can identify three charging regions for the \textcolor{black}{quantum} battery due to finite-size effects related to the spin chain~\cite{Rossini20} and we can plot the maximum energy stored per dimer as a function of $\delta_0$. The result for fixed $\delta_1 = 7$ is reported in Fig. \ref{SSH_Nearest} for the recurrence regime (red) and the asymptotic regime (green), i.e. $\tau \to \infty$. In the following plots, we define $\mathcal{N} = N/2$ as the number of dimers and $E_{max}$ as the maximum energy stored in the battery. Specifically, for the recurrence regime, $E_{max} = \Delta E^{SSH}(\tau_{max})$, representing the peak energy over time, while for the asymptotic regime, $E_{max} = E_{\infty}$. \textcolor{black}{Since this is, among all the analyzed models, the one that achieved the best performance in terms of the percentage of stored energy relative to the maximum capacity~\cite{Julia-Farre_2020, Hamma21}, a brief mention will also be made of the energy percentages associated with the various peaks.} As we can see, in the recurrence regime three peaks emerge corresponding respectively to $\delta_0 + \delta_1 = -1, 0$ and $1$, which are the values related to flat bands and the QPT line, with the highest one reaching approximately $60 \%$ of the total storable energy (not shown), while in the asymptotic regime only the peak related to the quantum phase transition ($\delta_0 + \delta_1 = 0$) remains, storing around $50 \%$ of the \textcolor{black}{maximum possible} energy. While Fig. \ref{SSH_NO_TEMP} shows the results at low temperature, Fig. \ref{SSH_TEMP} illustrates the effects of a high-temperature setting. In the latter, the peaks remain robust, with only a change in the percentage of energy stored by the device by approximately $10 \%$.
\subsubsection{First and second nearest-neighbor interactions}
According to Eq.~\eqref{d_vector_SSH}, adding a constant second nearest-neighbor interaction to the system does not affect $d_1(k,t)$ and $d_2(k,t)$, but introduces a term $d_0(k) = 2 J_2 \cos(k)$. In general, we observe that every even nearest-neighbor interaction contributes only to the diagonal terms of the Hamiltonian as part of the $d_0$ component. Another crucial point is that the only influence of $d_0$ on the formula for the stored energy, reported in Eq.\eqref{H_Gen_Non_Superconduttiva}, is through the temperature-dependent term, given by Eq.\eqref{FT}. Therefore, for $T = 0$, the formula for the energy stored does not take into account the $d_0$ component, while at non-zero temperatures we must account for the factor in Eq.\eqref{FT} that can be expressed as
\begin{equation}
    F_T(k,T,\mu) = \frac{\sinh(\beta \epsilon(k))}{\cosh(\beta \epsilon(k)) + \cosh(\beta(d^A_0(k) - \mu))}.
\end{equation}
\textcolor{black}{Indeed, this fact only shows that at finite temperature the position of the chemical potential with respect to the two levels defined at every $k$ does play a role.}
Notably, this factor does not affect the presence of energy storage peaks, leading us to conclude that, regardless of temperature, adding a second nearest-neighbor interaction does not alter the energy stored in the battery compared to the case with only first nearest-neighbor interactions, as long as for every quasimomentum only one electron is present. This result holds for any even nearest-neighbor interaction added to the system. To go deeper into our long-range analysis, we can see what happens in presence of both first and third nearest-neighbor interactions.
\subsubsection{First and third nearest-neighbor interactions}
\begin{figure}[H]
    \centering
    \includegraphics[width=0.9\linewidth]{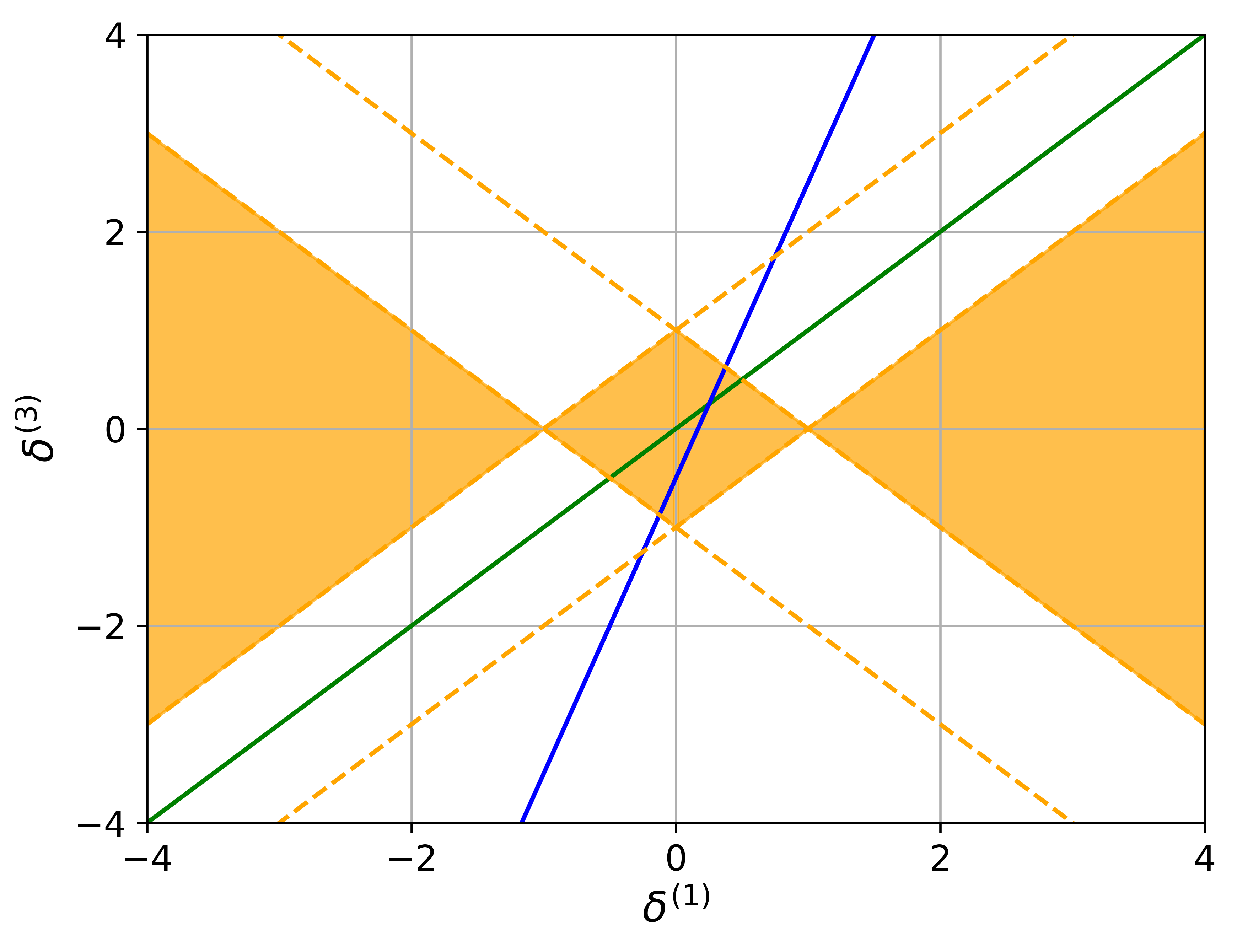}
    \caption{Physically-allowed regions (orange), where $J_1$ and $J'_1$ are greater in magnitude than $J_3$ and $J'_3$, in the $\delta^{(1)} - \delta^{(3)}$ plane for $r = 0$. The green line represents the QPT line $\delta^{(3)} = (\alpha / \beta)  \delta^{(1)}$ with $\alpha = \beta = 1$. The blue line $\delta^{(3)} = 3\delta^{(1)} - 0.5$ is the one on which the evolution of the system takes place.}
    \label{OrangeRegions}
\end{figure}

\begin{figure}[H]
    \centering
    \includegraphics[width=0.8\linewidth]{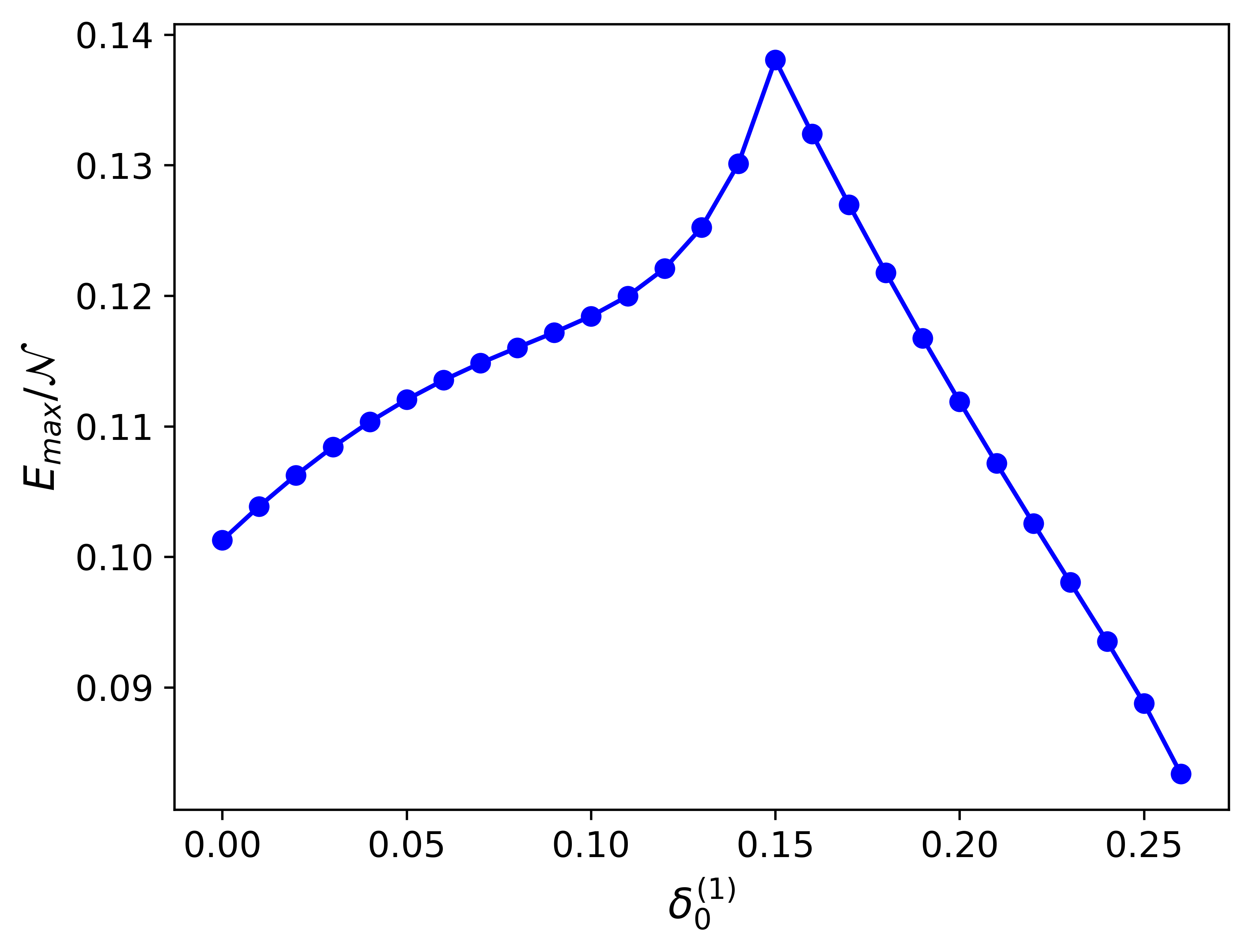}
    \caption{Maximum energy stored per dimer $\mathcal{N}=N/2$ in the $\tau \to \infty$ limit as a function of $\delta_{0}^{(1)}$ with $\delta_{1}^{(1)} = 0.1$, $\delta_0^{(3)} = 3\delta_0^{(1)} + 0.5$ and $\beta \to \infty$.}
    \label{SSH-Third}
\end{figure}

The only interactions that are relevant in the following analysis are $J_1, J'_1, J_3$ and $J'_3$. Now $p$ in Eq.\eqref{d_vector_SSH} can assume both $p=1$ and $p=2$ values, leading to
\begin{equation}
\begin{aligned}
&d_0(k,t) = 0, \\
&d_1(k,t) = \sum_{p=1,2} \left[J'_{2p-1}(t) \cos((p-1)k) + J_{2p-1}(t) \cos(pk)\right], \\
&d_2(k,t) = \sum_{p=1,2} \left[J_{2p-1}(t) \sin(pk) - J'_{2p-1}(t) \sin((p-1)k)\right]. \\
\end{aligned}
\end{equation}
In this new scenario, any charging protocol must account for the fact that in a realistic device, interactions between first and third nearest-neighbor sites are intrinsically linked: altering the former will inevitably affect the latter. For this reason, we decided to introduce two dimerization parameters, $\delta^{(1)}$ and $\delta^{(3)}$, respectively for first and third nearest-neighbor interactions, and evolve them in such a way that
\begin{equation}
    \delta^{(3)} = f(\delta^{(1)}) = m\delta^{(1)} + q
\end{equation}
with $m,q \in \mathds{R}$. So, we evolve both parameters from $(\delta^{(1)}_0, f(\delta^{(1)}_0))$ to $(\delta^{(1)}_0 + \delta^{(1)}_1, f(\delta^{(1)}_0+ \delta^{(1)}_1))$, where $\delta^{(1)}_1$ is the constant increment we give to the dimerization. It is also important to take into account the fact that, for our protocol to describe a physically realistic situation, the nearest-neighbor couplings must always be stronger in magnitude than those between the third nearest neighbors. Therefore, considering these conditions, the relations we choose between the $J$ parameters and the dimerization parameters are the following
\begin{equation*}
    \begin{cases}
        & J_1 = 1 - \alpha \delta^{(1)}, \quad J'_1 = 1 + \alpha \delta^{(1)} \\
        & J_3 = r - \beta \delta^{(3)}, \quad J'_3 = r + \beta \delta^{(3)}
    \end{cases}
\end{equation*}
where $\alpha$, $\beta$ and $r$ are real parameters tuned to satisfy the following conditions: (i) there must be at least one region in the $\delta^{(1)} - \delta^{(3)}$ plane where nearest-neighbor interactions are always stronger than third nearest-neighbor ones; (ii) this region, or at least one of them if there are more than one, must include a QPT in order to study its effect on the \textcolor{black}{QB}'s charging process. From Eq.\eqref{epsilon_non_sc}, trying to solve $\epsilon(k) = 0$ results in having complex values of $\delta^{(1)}$ except for two values of $k$: if $k = 0$ the system exhibits bands touching when $r = -1$ for every value of $\delta^{(1)}$ and $\delta^{(3)}$, while for $k = \pi$ bands touch for 
\begin{equation}
\delta^{(3)} = \frac{\alpha}{\beta} \delta^{(1)}. \label{QPTLine} 
\end{equation}
If we now impose $|J_1| > |J_3| ~ \bigwedge ~ |J_1| > |J'_3| ~ \bigwedge ~ |J'_1| > |J_3| ~ \bigwedge ~ |J'_1| > |J'_3|$ we notice that the only way to have a region that satisfies the aforementioned conditions and includes the QPT line in Eq.~\eqref{QPTLine} is by fixing $0 \leq r < 1$: in particular, the area of this region is maximized by setting $r = 0$. In Fig.~\ref{OrangeRegions} a sketch of the situation is represented: in the orange regions nearest-neighbor interactions are the strongest ones, while the green line represents the QPT line reported in Eq.\eqref{QPTLine} for $\alpha = \beta = 1$. It can be observed that the size of the diamond-shaped region in the middle of the plot can be changed by adjusting $\alpha$ and $\beta$: its diagonals along the $\delta^{(1)}$ and $\delta^{(3)}$ axes, in fact, are $1/\alpha$ and $1/\beta$ respectively. The blue line is the one we will use \textcolor{black}{for our charging protocol} and its equation is $\delta^{(3)} = 3\delta^{(1)} - 0.5$. According to the situation we have just described, if the presence of a QPT plays a role in the charging process of the \textcolor{black}{QB}, we should see something in $\delta^{(1)} = 0.25$, which is the value for which the green and blue lines coincide. 
\noindent In Fig.~\ref{SSH-Third} we can see the energy stored per dimer \textcolor{black}{for fixed $\delta_1^{(1)} = 0.1$}: the plot exhibits a distinguishable peak at $\delta_0^{(1)} = 0.15$. For this value of $\delta_0^{(1)}$, the couple of dimerization parameters that describes the charging Hamiltonian is
\begin{equation*}
    (\delta^{(1)}, \delta^{(3)}) = (\delta_0^{(1)} + \delta_1^{(1)}, 3 (\delta_0^{(1)} + \delta_1^{(1)}) - 0.5) = (0.25, 0.25)
\end{equation*}
showing, again, an enhancement of energy stored when the charging Hamiltonian is critical: this peak proves that \textcolor{black}{the observed} effects are robust \textcolor{black}{also} in presence of long-range hoppings, even though this scenario is energetically disadvantageous compared to the one with shorter-range interactions, as it only manages to store just over $5 \%$ of the \textcolor{black}{maximum possible} energy.
In all the regimes shown, it is worth to notice that the maximum charging power scales linearly in the number of lattice sites.


\section{Conclusions}
In this work, we investigated the dynamics of quantum batteries \textcolor{black}{based on} one-dimensional systems that can be \textcolor{black}{exactly solved through a mapping into} free fermions. By implementing a double quench protocol on a generic parameter in the battery's Hamiltonian, we derived a general analytical expression for the energy stored in the battery and we have applied this result across various \textcolor{black}{different} systems.

Applying the general results to the quantum Ising chain we showed that, even \textcolor{black}{at finite} temperature, the stored energy strongly depends, in a non-analytical fashion in the thermodynamic limit, on the presence of a QPT at $h = 1$. Moreover, while increasing the temperature drastically reduces the stored energy, we identified the formation of a plateau just before reaching the critical value of the external field. By altering the charging protocol and initializing the process with the external field set to zero, we demonstrated that this plateau can be recovered even at zero temperature.

Further applying the formula to the more complex XY model, we found that the plateau also emerges in this model, again provided the process starts from the classical point of the model, which in this case is $(\gamma_0, h_0) = (1,0)$, with a quench in the anisotropy parameter.

Furthermore, to investigate the role of longer-range terms, we examined the cluster Ising model. Although we do find strong signatures of the quantum phase transitions between phases dominated by different ranges of the interactions, the charging power scales linearly with the number of sites.

Finally, we studied the extended SSH model, revealing a strong dependence of the stored energy on the quantum phase diagram, indicated by the presence of peaks. However, even in the presence of the longer range hoppings, just as in the cluster Ising model, the charging power is linear in the number of lattice sites.

\textcolor{black}{Future developments of the present work will involve the investigation of different charging protocols and the study of the robustness of the results achieved with respect to the coupling of the considered models with a dissipative environment.}

\section*{Acknowledgments}
  N.T.Z. acknowledges the funding through the NextGenerationEu Curiosity Driven
Project "Understanding even-odd criticality". N.T.Z. and M.S. acknowledge the funding
through the "Non-reciprocal supercurrent and topological transitions in hybrid Nb- InSb
nanoflags" project (Prot. 2022PH852L) in the framework of PRIN 2022 initiative of the
Italian Ministry of University (MUR) for the National Research Program (PNR). This project has been funded within the programme ``PNRR Missione 4 - Componente 2 - Investimento 1.1 Fondo per il Programma Nazionale di Ricerca e Progetti di Rilevante Interesse Nazionale (PRIN)''

D.F. acknowledges the contribution
of the European Union-NextGenerationEU through the
"Quantum Busses for Coherent Energy Transfer" (QUBERT) project, in the framework of the Curiosity Driven
2021 initiative of the University of Genova and through
the "Solid State Quantum Batteries: Characterization
and Optimization" (SoS-QuBa) project (Prot. 2022XK5CPX), in the framework of the PRIN 2022 initiative of the Italian Ministry
of University (MUR) for the National Research Program
(PNR). This project has been funded within the programme ``PNRR Missione 4 - Componente 2 - Investimento 1.1 Fondo per il Programma Nazionale di Ricerca e Progetti di Rilevante Interesse Nazionale (PRIN)''.

\section*{Competing interests}
All authors declare no competing interests.

\appendix
\section{Explicit calculation of the integral in Eq.\eqref{integrale_xy}}\label{Appendix}
In this appendix we will derive the explicit analytical computation of the integral in Eq.\eqref{integrale_xy}. The indefinite integral is
\begin{equation}
    \int \frac{\sin^4(k) - \sin^2(k)}{1 + (\gamma_1^2 - 1)\sin^2(k)} dk.
\end{equation}
After multiplying both numerator and denominator for $(\gamma_1^2 - 1)^2$, we have
\begin{equation}
    \int \frac{(\gamma_1^2 - 1)^2\sin^4(k) - (\gamma_1^2 - 1)^2\sin^2(k)}{(\gamma_1^2 - 1)^2[1 + (\gamma_1^2 - 1)\sin^2(k)]} dk. \label{int_app_A2}
\end{equation}
Now the numerator can be factorized as follows
\begin{equation}
    (\gamma_1^2 - 1)^2\sin^4(k) - (\gamma_1^2 - 1)^2\sin^2(k) = \gamma_1^2 + [1 + (\gamma_1^2 - 1)\sin^2(k)][-\gamma_1^2 + (\gamma_1^2 - 1)\sin^2(k)]
\end{equation}
and this allows the integral in Eq.\eqref{int_app_A2} to be written as
\begin{equation}
    \int \frac{\gamma_1^2}{(\gamma_1^2 - 1)^2[1 + (\gamma_1^2 - 1)\sin^2(k)]} dk + \int \frac{-\gamma_1^2 + (\gamma_1^2 - 1)\sin^2(k)}{(\gamma_1^2 - 1)^2} dk. \label{int_sum}
\end{equation}
The first integral in Eq.\eqref{int_sum} can be written as
\begin{equation}
    \frac{\gamma_1^2}{(\gamma_1^2 - 1)^2} \int \frac{1}{1 + (\gamma_1^2 - 1)\sin^2(k)} dk
\end{equation}
that can be solved by means of the substitution $y \equiv \tan(k)$. The integral becomes
\begin{equation}
    \frac{\gamma_1^2}{(\gamma_1^2 - 1)^2} \int \frac{1}{1 + (\gamma_1 y)^2} dk = \frac{\gamma_1^2}{(\gamma_1^2 - 1)^2} \frac{\arctan(\gamma_1 y)}{\gamma_1} = \frac{\gamma_1}{(\gamma_1^2 - 1)^2} \arctan(\gamma_1 \tan(k)). \label{result_1}
\end{equation}
The second integral of Eq.\eqref{int_sum} can be solved using linearity
\begin{equation}
    \int \frac{-\gamma_1^2 + (\gamma_1^2 - 1)\sin^2(k)}{(\gamma_1^2 - 1)^2} dk = -\frac{\gamma_1^2}{(\gamma_1^2 - 1)^2} \int dk + \frac{1}{\gamma_1^2 - 1} \int \sin^2(k) dk
\end{equation}
and the result is
\begin{equation}
    -\frac{\gamma_1^2}{(\gamma_1^2 - 1)^2} k + \frac{1}{\gamma_1^2 - 1} \left(\frac{k}{2} - \frac{\sin(2k)}{4}\right). \label{result_2}
\end{equation}
So, the complete result of the indefinite integral is the sum of the results reported in Eq.\eqref{result_1} and Eq.\eqref{result_2} respectively. When we evaluate this result in the Brillouin zone, we have to take into account the fact that the term reported in Eq.\eqref{result_1} is not a continuous function of $k$ in the interval $\left[-\pi,\pi\right]$, so it must be treated from $0$ to $(\pi/2)^-$ and from $(\pi/2)^+$ to $\pi$ separately. \textcolor{black}{By carefully keeping track of this technical issues}, the integral in Eq.\eqref{integrale_xy} gives the result reported in Eq.\eqref{Result_Integral_XY} of the main text.

\bibliographystyle{elsarticle-harv}
\bibliography{elsarticle-template-harv}

\end{document}